\documentstyle[prb,preprint,aps]{revtex}
\draft
\begin{document}
\title{A HIGH-RESOLUTION COMPTON SCATTERING STUDY OF THE ELECTRON MOMENTUM 
DENSITY IN Al}
\author{T. Ohata, M. Itou}
\address {Japan Synchrotron Radiation Research Institute, Kamigouri,
Akougun, Hyogo 678-12, Japan}
\author{I. Matsumoto}
\address {Department of Synchrotron Radiation Science, Graduate
University for Advanced Studies, Tsukuba 305, Japan}
\author{Y. Sakurai}
\address {Japan Synchrotron Radiation Research Institute, Kamigouri,
Akougun, Hyogo 678-12, Japan}
\author{H. Kawata}
\address {Institute of Materials Structure Science, High Energy
Accelerator Research Organization, Tsukuba, Ibaraki 305, Japan}
\author{N. Shiotani}
\address {Tokyo University of Fisheries, Kounan, Minato, Tokyo 108, Japan}
\author{S. Kaprzyk}
\address {Academy of Mining and Metallurgy, Al.Mickiewicza 30, Krakow, Poland  
\\and Department of Physics, Northeastern University, Boston, MA 02115, USA}
\author{P.E. Mijnarends}
\address {Department of Physics, Northeastern University, Boston, MA 02115, USA
\\and Interfaculty Reactor Institute, Delft University of Technology,
\\ 2629 JB Delft, The Netherlands}
\author{A. Bansil}
\address {Department of Physics, Northeastern University,
                                   Boston, MA 02115, USA}
\date{\today}
\maketitle
\begin{abstract}

We report high-resolution Compton profiles (CP's) of Al along the 
three principal symmetry directions at a photon energy of 59.38 keV, 
together with corresponding highly accurate theoretical profiles 
obtained within the local-density approximation (LDA) based band-theory 
framework. A good accord between theory and experiment is found with 
respect to the overall shapes of the CP's, their first and second 
derivatives, as well as the anisotropies in the CP's defined as differences 
between pairs of various CP's. There are however discrepancies 
in that, in comparison to the LDA predictions, the measured profiles are 
lower at low momenta, show a Fermi cutoff which is broader, and display a 
tail which is higher at momenta above the Fermi momentum. A number of 
simple model calculations are carried out in order to gain insight into 
the nature of the underlying 3D momentum density in Al, and the role of 
the Fermi surface in inducing fine structure in the CP's. The present 
results when compared with those on Li show clearly that the size of 
discrepancies between theoretical and experimental CP's is markedly 
smaller in Al than in Li. This indicates that, with increasing electron 
density, the conventional picture of the electron gas becomes more 
representative of the momentum density and that
shortcomings of the LDA framework in describing the electron correlation
effects become less important.  \\

\end{abstract}

\section{INTRODUCTION}

In a Compton scattering experiment one measures the
so-called Compton profile (CP),
\begin{equation}
              J(p_z)=\int \int \rho({\mathbf p}) dp_x dp_y,
\label{equ:01}
\end{equation}
where $\rho({\mathbf p})$ is the ground state electron momentum density.
In an independent particle model the momentum density is given by
\begin{equation}
    \rho({\mathbf p}) = (2\pi)^{-3} \sum | \int \psi({\mathbf r})
                       \exp(i{\mathbf p} \cdot {\mathbf r}) d{\mathbf r}|^2,
\label{equ:02}
\end{equation}
where $\psi({\mathbf r})$ denotes the electron wave
function.\cite{1,bansil93,mijn,manni00}
The summation
in Eq.~(\ref{equ:02}) extends over all occupied states.  The Compton profile,
$J(p_z)$, thus contains signatures of the Fermi surface
breaks and correlation effects in the underlying three
dimensional momentum distribution $\rho({\mathbf p})$.  Since Fermi momenta
$p_f$ are typically $\sim$ 1 a.u., a high momentum
resolution of $\sim$0.1 a.u. is essential in the experiment for
delineating Fermi surface related fine structure in the CP. 

High-resolution Compton studies have recently been reported on
Li,\cite{saku95,schulke96} Be,\cite{hama96,itou98,huot00} 
V,\cite{shio93} and Cu.\cite{saku99}
In all these cases, careful comparisons of the shapes of the 
absolute valence electron CP's, and the structure in the first and 
second derivatives as well as the directional anisotropies of the CP's, 
have been made with the corresponding parameter-free 
theoretical predictions based on the use of the LDA. 
A similar investigation of Li-rich LiMg disordered alloys where 
disorder effects were treated using the mean field KKR-CPA approach has
also been carried out.\cite{stutz99}  In this way, the band-theory based 
LDA approach has been shown to provide a remarkably accurate description 
of many aspects of the momentum density associated with 
the quantum mechanical electronic ground state, 
including the characteristic fine structure induced by the Fermi surface.
More exciting however is the fact that the aforementioned 
comparisons have for the first time clearly established
the presence of systematic deviations 
between theoretical and experimental momentum densities.
In Li, the experimental break $Z_k$ in the momentum density at $p_f$ 
appears to be very small, nearly zero\cite{schulke96};
if so, this is very far from 
the results of electron gas calculations stretching over the last several 
decades.\cite{foot1}
In Be, the latest Compton data\cite{huot00} indicates {\em anisotropic} 
electron correlation effects outside the scope of much of the existing 
theoretical work which is based on treating properties of the 
homogeneous electron
gas.\cite{daniel60,lund67,over71,lantto80,farid93,holm98}
For these reasons, a renewed interest in the 
problem of correlation effects on the momentum density beyond the 
LDA is natural,\cite{kralik98,filippi99,barb00,eguiluz00,kubo96} 
although much further work is necessary for developing 
an approach of wide applicability in metals and alloys.

Bearing these considerations in mind, there is strong motivation for 
undertaking high-resolution Compton studies of other systems. Our 
choice of Al in this connection is an especially good one
because Al is trivalent and, therefore, it extends 
the range of electron densities investigated so far via high-resolution
Compton. Correlation effects are of course expected to become less important 
with increasing electron density as the kinetic energy dominates. 
Also, Al has been the traditional touchstone of a 
free-electron-like metal with a nearly spherical Fermi surface 
(viewed in the extended zone). Neither a high-resolution, high-statistics
Compton measurement, nor a band theory computation of high accuracy 
in order to identify Fermi surface related fine structure in the CP's of Al 
is currently available in the literature.\cite{foot2,suort00}
The goal of the present
work is to fill this gap and determine the extent to which the LDA 
describes the momentum density in Al. The existing Compton 
data on Al consists essentially of a number of measured CP's using 
$\gamma$-ray sources and solid state detectors at low 
momentum resolution.\cite{8,9,10,11} Quite some time ago, 
Shiotani {\em et al.}\cite{12} obtained the [111] CP of Al at a momentum 
resolution of 0.08 au, but did not investigate the anisotropy of the 
CP or the Fermi surface signatures therein. 

An outline of this article is as follows. 
In the next section we describe the experimental
procedures. Section III gives pertinent details of 
computations. In section IV the experimental CP's are analyzed 
in the light of the band theory predictions as well as a number of other 
model computations. The Compton results are also compared briefly with
closely related positron annihilation spectra. Section V summarizes our
main conclusions.
\\

\section{EXPERIMENT}

Single crystals of Al with surface normals oriented along
the [100], [110] and [111] directions were used. The thickness
of the crystals was about 2 mm. The reader is referred to
Sakurai {\em et al.}\cite{13} for details of our Compton spectrometer,
and to Tanaka {\em et al.}\cite{14} and Itou {\em et
al.}\cite{itou98} for our data processing procedures. Briefly, the spectrometer
consists of a Cauchois-type bent-crystal analyzer of Si(422)
with an image plate serving as a position sensitive detector.
The scattering angle is $160^{\circ}$. The synchrotron radiation
X-rays from a multipole wiggler installed in the 6.5 GeV
Accumulation Ring at the National Laboratory for High Energy
Physics are monochromatized by a quasi-doubly bent
monochromator to 59.38 keV with an energy resolution of about
80 eV. The overall momentum resolution is estimated to be 0.12 a.u.  The
double Compton scattering events were simulated via the Monte
Carlo program of Sakai\cite{15}; the integrated intensity of
the double scattering events was found to be 10\% of the single
scattering events. 

The statistical error of each datum
point, given by $\sigma = \sqrt N + 0.003 N$, is estimated to be less
than 0.3 \%. Since the data points are not measured equidistantly they
are interpolated onto an equidistant mesh of 0.02 a.u. using simple
linear interpolation. The data were
numerically differentiated according to
$y'(i) = \frac{1}{2}\lbrace [y(i+1)-y(i)]/[x(i+1)-x(i)] + 
[y(i)-y(i-1)]/[x(i)-x(i-1)]\rbrace$. No further smoothing or filtering
was applied. The interpolation and differentiation cause some 
statistical correlation between the data points. \\

\section{COMPUTATIONS}

The band structure problem was solved within the 
all-electron charge self-consistent KKR framework without any free
parameter. Exchange-correlation effects were
incorporated using the von Barth-Hedin local spin density
(LSD) approximation.\cite{16}  The lattice constant was computed to
be 7.6534 a.u. by minimizing the total energy; for comparison,
the experimental lattice constant at room temperature is 7.6559
a.u. The self-consistent
crystal potential was obtained by iterating the KKR cycles
using an elliptic contour with 48 points in the complex energy
plane. The final charge density is self-consistent to an
accuracy of about $10^{-4}$ electrons and the Fermi energy to $10^{-4}$ Ry. 
An angular momentum cutoff $l_{max}=2$ was employed.
A free-electron-like Fermi surface was found with Fermi radii
$k_{100} = 0.9246, k_{110} = 0.9255,$ and $k_{111} = 0.9292$ a.u.;
the free-electron value would be 0.9254 a.u.

The CP's were obtained by first evaluating the three-dimensional
momentum density $\rho({\mathbf p})$ in terms of the
momentum matrix element of the KKR Green's function\cite{17,18,19,ban99}
over a fine mesh of $48 \times 4851 \times 177$ {\bf p}-points,
covering momenta up
to $p_{max} \sim 5$ a.u. This mesh involves 4851 {\bf k}-points in the
1/48-th irreducible part of the Brillouin zone with each
{\bf k}-point translated into 177 {\bf p}-points by adding reciprocal
lattice vectors; the factor of 48 takes into account the symmetry
operations of the cubic point group.
The CP's can then be computed accurately by
evaluating the two-dimensional integral of Eq.~(\ref{equ:01}) using a
generalized linear tetrahedron method.\cite{19a}
The final CP's
have been calculated over a momentum mesh containing
151 $p_z$ points in the range 0-3 a.u. along each of the three
measured directions. The accuracy of the computed profiles is
about 1 part in $10^4$. A similar integration technique has been 
used in our earlier studies of high-resolution CP's of various metals and 
alloys.\cite{saku95,hama96,itou98,huot00,saku99,suort00,CuNi,matsu00}. 
The Lam-Platzman correction\cite{lam74} to the CP's was computed using the 
occupation number density of the uniform electron gas. 
\\

\section{RESULTS AND DISCUSSION}

Figure~\ref{fig1} shows the measured and computed CP's of
the valence electrons along the [100], [110] and [111]
directions; the theoretical CP's are convoluted with a gaussian
which represents the experimental resolution of 0.12 a.u. FWHM.
The experimental valence CP's have been obtained by
subtracting the theoretical core CP's from the measured profiles.  
In this connection, we used the solid-state core wavefunctions which 
reflect the slight overlap of the $2p$ core states in Al. 
The first and second derivatives of the valence profiles have been
obtained by numerical differentiation.

In examining the overall shape of the CP's in Fig.~\ref{fig1}, 
one notes that the experimental points are lower 
at low momenta compared to the calculated values. We emphasize that this 
does not imply that the measured 3D momentum density  is lower than the 
theoretical one at all momenta. 
To see this, recall that\cite{m69}
\begin{equation}
  \rho(0) = \left . -\frac{1}{2\pi} \frac{d^2J_{av.}(p)}{dp^2} \right|_{p=0},
\label{equ:03}
\end{equation}
where $J_{av.}(p)$ denotes the directionally averaged CP which in a cubic 
crystal may be reasonably approximated by\cite{miasek}
\begin{equation}
    J_{av.}(p) = (1/35)[10 J_{100}(p) + 16 J_{110}(p)+ 9 J_{111}(p)].
\label{equ:04}
\end{equation}
The bottom row in Fig.~\ref{fig1} shows that the differences between the 
experimental and theoretical second derivatives at $p_z=0$ are well within 
the error bars. In view of Eq. (\ref{equ:03}), this indicates that the
underlying 3D distributions are not significantly different at $p_z=0$.
In fact, this result implies that the measured momentum density
must be smaller than the theoretical one at momenta
approaching the Fermi momentum $p_f$. This is also borne out by
the first derivatives shown in the central row of Fig.~\ref{fig1} which
begin to show differences between experiment and theory only
above $p_z \sim 0.3$ a.u. 

Further insight is provided by Fig. \ref{fig2} which shows the spherically 
averaged 3D momentum density defined by 
\begin{equation}
       \rho_{\mbox{\scriptsize av.}}(p) = -(1/2 \pi p)(dJ_{av.}/dp),
\label{equ:05}
\end{equation}
where $J_{\mbox{\scriptsize av.}}(p)$ is obtained via Eq. (\ref{equ:04}). 
The oscillations in $\rho_{\mbox{\scriptsize
av.}}(p)$ at small momenta reflect partly the large (correlated)
error bars due to the division of the small derivative by small
values of $p$, and partly the (spherically averaged) effect of
Brillouin zone-face interactions to be discussed below.
In any event, Fig. \ref{fig2} makes it clear that the experimental 3D momentum 
density lies below the theoretical predictions as one approaches
$p_f$, and that the situation reverses itself above $p_f$.

The 2nd derivatives $J^{\prime \prime}$ in Fig. \ref{fig1} all show
a peak at $p_f$.
There is good agreement between theory and experiment as to the
position of the peaks, {\em i.e.}, the value of $p_f$, but the measured
peaks are all lower and broader than the theoretical predictions.
Although the shapes of the $J^{\prime \prime}$ peaks
reflect the complex interplay
between the effects of experimental resolution, electron correlations and
lattice potential on the Fermi cut-off in the momentum density,
it is evident from Figs. \ref{fig1} and \ref{fig2} that the measured
distribution possesses a tail higher than the theory beyond $p_f$.

These discrepancies between theory and experiment
are similar to those reported earlier in 
Li\cite{saku95,schulke96} and other metals\cite{hama96,itou98,huot00,saku99} 
and have their origin in the electron 
correlation effects beyond the LDA which are not treated properly 
within our theoretical framework. 
Such correlations are expected to 
cause (relative to the independent particle model)
a decrease of the momentum density as one approaches $p_f$, 
and a tail at momenta greater than $p_f$; as indicated above, both
features are qualitatively visible
in our comparison between theory and experiment. 
Notably, the deviations
from LDA theory are smaller in Al
than in Li. For example,
the difference between the theoretical and experimental valence profiles
at $p_z=0$ is approximately 16 \% for Li and 4.5 \% for
Al,\cite{foot3} and the width
of the peak at $p_f$ in the second derivatives
is 0.23 a.u. in Li and 0.15
a.u. in Al; thus, the "blurring" of the Fermi cutoff is more severe in
Li than in Al. These characteristic differences between Li and Al
are partly related to the difference in the electron density of
the two metals.  The electron density in terms of $r_s$ (the
standard parameter for the volume per electron of valence electrons), 
is 3.21 for Li and 2.12 for Al. Therefore, the bare
Coulomb interaction is more effectively screened in Al than in Li.
As shown by a variety of treatments of the homogenous interacting electron 
gas, as the electron density increases, the kinetic energy dominates, and 
the momentum density is described more closely by the
free-electron rectangular distribution with a step-wise cutoff at
$p_f$.\cite{daniel60,lund67,over71,lantto80,farid93,holm98,lam71,cep80,foot4}

Figure 3 considers the effect of the isotropic Lam-Platzman (LP) correction 
on the [111] CP; results along other directions are similar and are not shown 
in the interest of brevity. The theoretical curves in Fig. \ref{fig3}
include the LP-correction, while those in Fig. \ref{fig1} do not.
A comparison of Fig. \ref{fig3} with 
the last column of Fig. \ref{fig1} shows that, although the inclusion of the
LP-correction improves things, much of the discrepancy between
theory and experiment still remains. Interestingly, Ref. \onlinecite{suort00}
has recently analyzed the correlation correction to the CP's of Al in terms
of a model which involves the break $Z_k$ in the momentum density at
$p_f$ as the only free parameter. By adjusting $Z_k$,
Ref. \onlinecite{suort00} finds that the discrepancy between the
LDA predictions and the measurements can be essentially removed for a
$Z_k$ value between 0.7 and 0.8, in reasonable accord with
the corresponding theoretical values from various authors which are
scattered between 0.76 and 
0.85.\cite{daniel60,lund67,over71,lantto80,farid93,holm98}
There is no inconsistency between the present results
and those of Ref. \onlinecite{suort00}. To see this relationship,
recall that the standard LP-correction is defined via\cite{lam74}
\begin{equation}
 \Delta\rho(p)=\int d^{3}r\rho({\mathbf{r}})
 \left[
  \rho^{INT}(p,r_{s}({\mathbf{r}}))
 -\rho^{NI}(p,r_{s}({\mathbf{r}}))
 \right],
\label{equ:06}
\end{equation}
where the integral extends over the Wigner-Seitz cell. The expression within
the square brackets gives the difference between
the momentum densities of the interacting
and non-interacting {\em homogeneous} electron gas (denoted by the superscripts
`INT' and `NI') evaluated at the local density
$\rho({\mathbf{r}})$ of the physical system and $r_{s}({\mathbf{r}})$ is the
corresponding electron density parameter. Equation (\ref{equ:06}) thus
attempts to take into account
inhomogeneities in the electron gas, whereas the semi-empirical model of
Ref. \onlinecite{suort00} replaces the integrand by its value at
the average electron density in Al. The matter is quite subtle, and
further work is necessary in order to develop a satisfactory treatment of
correlation effects on the momentum density in solids.

If the momentum density within the Fermi sphere were flat and
smooth, the first derivative $dJ(p_z)/dp_z$ shown in the middle
row of Fig.~\ref{fig1} would be a straight
line up to the cutoff at the Fermi momentum. However, at momenta
less than the Fermi radius some structure is visible notably in the 
[111] and the [100] derivatives.
In this connection, we note that the Fermi sphere overlaps with 
Umklapp Fermi spheres centered on the (111) reciprocal lattice points
around the W-points in the Brillouin zone. 
For example, the hexagonal zone face contains six W-points 
which all project at ($\frac{1}{2},\frac{1}{2},
\frac{1}{2}$), {\em i.e.}, the point $p_z = 0.71$ a.u. on the [111]
axis. Similarly, four W points in the first Brillouin zone
project at 0.41 a.u. and another four at 0.82 a.u. on the [100] direction. 
Interestingly, the experimental as well as the theoretical
derivatives contain structure around 0.7 a.u. in the [111] and 0.4 a.u. 
in the [100] CP. 
 This indicates the importance 
of the k states near the W points with respect to the fine 
structure in the Al CP's. Incidentally, a structure similar to 
the wiggle around 0.4 a.u. in the [100] derivative has been 
observed in positron annihilation 1D-ACAR measurements by Okada 
{\em et al.}\cite{22} and 2D-ACAR measurements by Mader
{\em et al.}\cite{mad} who also 
ascribed it to zone-face interactions around the W-points.

The directional differences, shown in Fig.~\ref{fig4}, are a measure of
the anisotropy. Although the maximum difference is about 1 \% of
the peak value of the profile itself, they show definite
structures which can have several origins. Firstly, for different
crystal orientations the plane of integration in Eq.~(\ref{equ:01})
sweeps differently through the Umklapp Fermi spheres centered at
the reciprocal lattice points in the higher Brillouin
zones. Secondly, the Fermi surface is slightly
distorted from a sphere, as witnessed by the different Fermi
radii given above, while, thirdly, band-structure effects such as a
{\bf p} dependence of the momentum density within the Fermi spheres
and interactions of the electron bands
with the Brillouin zone faces with consequent distortion of the
wavefunctions will also contribute to the anisotropy. The
importance of the first point can be readily studied using a
simple model of a spherical free-electron-like Fermi surface
surrounded by seven shells of similar Umklapp Fermi surfaces. The
momentum density within each Fermi surface is assumed to be
constant and given by the square of the corresponding Fourier
component of the electron wave function at $\Gamma_1$.\cite{mijn}
The CP for a given direction then consists of a
superposition of inverted parabolas, centered at the projections on that
direction of the reciprocal lattice points.
The height of each parabola is proportional
to the momentum density within the corresponding Fermi sphere, while
its cut-off points are found by adding or subtracting $p_f$ from the
projected center. Figure~\ref{fig5} shows the directional differences thus
obtained. The positions of the cut-off points have been indicated by
the arrows at the bottom of the graph, together with a symbol which
denotes the direction of projection ($a=[100]$, $b=[110]$, $c=[111]$) and
the coordinates of the center of the Fermi sphere. It should be noted
that many Umklapp Fermi spheres coincide in projection and therefore
these coordinates are not unique; the simplest set has been
noted. The analysis in Fig.~\ref{fig5} shows that much of the important
structure in the directional differences stems from the $<111>$ Umklapp
contributions; the other Umklapps play a less important role.

A comparison of Fig.~\ref{fig5} with the calculated differences
in Fig.~\ref{fig4} shows an
overall qualitative correspondence in the succession of positive
and negative peaks. On a more detailed scale, however, there are
significant differences which have their origin in the other
factors mentioned above. Notable examples are the peaks around
0.85 a.u. in the $J_{[111]} - J_{[100]}$ and $J_{[111]} - J_{[110]}$
directional differences in Fig.~\ref{fig4} which have no clear counterpart
in Fig.~\ref{fig5}. Kubo {\em et al.}\cite{21} have ascribed these
features to the fact that in the $[111]$ direction the actual
Fermi surface bulges out beyond the free-electron
Fermi sphere in the second Brillouin zone while there is a contraction
in the third zone. This will strongly affect the $[111]$ profile but
not so much the other two. Our simple free-electron model of course
does not contain this Fermi surface distortion effect.
The calculated curves in Fig.~\ref{fig4},
on the other hand, include all of these factors and reproduce the
essential characteristics of the measured differences, although
some discrepancies remain. It may be noted that non-locality of the
exchange and correlation potential in Li reduces the Fermi
surface anisotropy\cite{rasolt,macdonald} and thus would affect
the anisotropy of the CP's. In this vein, lattice vibrations
would reduce the Umklapp contributions and hence the anisotropy
of the momentum density. In how far such effects can explain the
residual discrepancies in Fig.~\ref{fig4} remains unclear.

In principle, the anisotropy in the momentum density may be
obtained approximately by expanding both the momentum density and
the CP's into lattice harmonics, and establishing the
relation between the expansion coefficients for the momentum
density and those for the CP's.\cite{m67} Actually,
Eq.~(\ref{equ:05}) represents the $l=0$ term in such a scheme.
However, we have not attempted to analyze our data along these lines
since the number of measured profiles is not large enough.

Additional information may be gained from a comparison of
CP's with the corresponding results of
positron annihilation measurements. Both experiments probe the 
momentum density -- 
in positron annihilation one
measures the momentum density of the annihilating 
electron-positron pair whereas
in Compton only the 
electron momentum density is involved.
In Fig.~\ref{fig6} the first derivative of the one-dimensional
angular correlation of positron annihilation radiation (1D-ACAR) profile 
measured by Okada {\em et al.}\cite{22} for the [111] orientation
is compared with the
corresponding CP. The momentum resolution of the 1D-ACAR
is 0.11 a.u. which is almost the same as that of the present
CP's. Since the 1D-ACAR was area normalized to the
theoretical 1D-ACAR calculated by Kubo {\em et al.},\cite{21} 
the peak height at $p_z=0$ is almost the same as that of
the present CP. The slope at the Fermi momentum
is steeper in the 1D-ACAR than in the CP, which is direct
evidence for enhancement of the annihilation of positrons
with the $s-p$ electrons near the Fermi energy predicted first by
Kahana\cite{kahana}
on the basis of an interacting electron gas model.
Also, the correlation tail for $p > p_f$ in the 1D-ACAR is weaker
than its counterpart in the CP as a result of the partial cancellation
of electron-electron and positron-electron
correlation effects.\cite{carbotte}
Finally, the fine structure at 0.2 a.u. and 0.7 a.u. is more
pronounced in the 1D-ACAR than in the CP. This
points to less correlation-induced smearing in positron annihilation
compared to Compton scattering.
\\

\section{SUMMARY AND CONCLUSIONS}

We have measured the Compton profiles (CP's) of Al along [100], 
[110] and [111] directions at a photon energy of 59.38 keV and a 
momentum resolution of 0.12 au. Parallel, highly accurate all-electron
computations have been carried out within the LDA-based band-theory
framework. Comparisons between theory and experiment at the level of 
the shapes of the CP's, structure in the 1st and 2nd derivatives of the 
CP's, and anisotropies obtained by taking differences between 
three pairs of CP's, all show a good level of accord. 
However, there are discrepancies as well. In comparison to the 
LDA predictions, the measured profiles are lower at low momenta, 
show a Fermi cutoff which is broader, and display a 
tail which is higher at momenta above the Fermi momentum.
The inclusion of correlation effects in the LDA via the standard
isotropic Lam-Platzman correction improves things slightly, but the 
essential discrepancies remain. A model analysis in 
terms of directionally averaged CP's allows us to get
a handle on the 3D momentum density of Al; in this way, we adduce that 
the experimental 3D density near p=0 does not differ significantly
from LDA predictions even though the CP's do. 
In this vein, CP's are computed using a model 3D distribution in which 
free electron spheres with appropriate weights are placed on reciprocal 
lattice points (extending to seven shells around a central sphere) to
represent the higher momentum components in the electronic wave functions; 
the results show that a significant amount of fine structure in the CP's 
is induced by these higher momentum components and by {\bf k}-states
near the W-points in the Brillouin zone where
the free electron spheres overlap. 

The present results when compared with those reported earlier
on Li show clearly that the size of discrepancies between theoretical and 
experimental CP's is markedly smaller in Al than in Li; in particular, 
theoretical and experimental profiles at $p_z=0$ differ by 16$\%$
in Li but only by 4.5$\%$ in Al, and the peak width in the 2nd derivative
at $p_f$ is 0.23 au in Li but 0.15 au in Al. 
It is thus clear that, with increasing electron density, 
the conventional picture of the electron gas becomes more 
representative of the momentum density and that shortcomings of the LDA
framework in describing the electron correlation 
effects become less important.
Finally, we compare briefly our [111] CP with the positron-annihilation
(1D-ACAR) measurements of Okada et al., and show that in the case of 
positron-annihilation the Fermi cut-off is 
sharper and that there is less correlation 
induced smearing of structures in the ACAR spectrum. 
\\

\acknowledgments

It is a pleasure to acknowledge important conversations with Bernardo
Barbiellini. The Compton profile measurements were performed with the
approval of the Photon Factory Advisory Committee, Proposal
Nos. 92-G257, 94-G351 and 97-G288.
This work is supported by the US Department of Energy under contract
W-31-109-ENG-38, by the Polish Committee for Scientific Research
through grant number 2P03B02814, and benefited from a travel grant 
from NATO, and the allocation of supercomputer
time at NERSC and the
Northeastern University Advanced Scientific Computation Center (NU-ASCC).



\begin{figure}
\caption{
Top: Measured and computed Compton profiles of Al along
the [100], [110] and [111] directions. Theoretical
profiles (solid lines) have been broadened to reflect
experimental resolution. Middle: First derivatives of the
measured and computed profiles. Bottom: Second derivatives
of the measured and the computed profiles.}
\label{fig1}
\end{figure}

\begin{figure}
\caption{
Theoretical (solid curve) and experimental (dashed curve)
directionally averaged 3D electron momentum density obtained via
Eqs. (\ref{equ:04}) and (\ref{equ:05}).}
\label{fig2}
\end{figure}

\begin{figure}
\caption{
Same as the last column of Fig. \ref{fig1}, except that here
the theoretical curves
in all cases include the Lam-Platzman correction.}
\label{fig3}
\end{figure} 

\begin{figure}
\caption{
Measured and computed directional difference profiles 
for three different pairs of directions.}
\label{fig4}
\end{figure}

\begin{figure}
\caption{Directional difference profiles 
calculated for a simple
quasi-free-electron model of Al in which the CP is given by a superposition 
of parabolic contributions centered at various reciprocal lattice points 
(see text). The arrows at the bottom
indicate the positions of the cut-off points (each parabola has two
cut-off points; the other one lies outside the graph). The directions
of projection are indicated by $a$ (=[100]), $b$ (=[110]), and $c$ (=[111]),
while the subscripts denote the coordinates of the reciprocal lattice
points involved. 000 denotes
the cut-off of the central Fermi surface, {\em i.e.}, the Fermi radius $p_f$.}
\label{fig5}
\end{figure}

\begin{figure}
\caption{First derivative of the 1D-ACAR spectrum (open circles) along the
[111] direction read off from Ref.\protect \onlinecite{22} is compared
with the derivative of the [111] Compton profile shown in
Fig. \protect \ref{fig1}.}
\label{fig6}
\end{figure}


\begin{references}
\bibitem{foot1}
  We will not concern ourselves here with questions related to 
  uncertainties of background and core contribution in 
  obtaining valence profiles from the Compton data, 
  or possible breakdown of the impulse approximation at low photon 
  energies.
\bibitem{foot4}
  The difference between Al and Li based on the 
  interacting gas computations is however relatively small. For example, 
  the value of the break $Z_k$ in the momentum density at $p_f$ 
  in going from Al to Li is estimated to decrease by only
  $\sim 10 \%$ which would not explain the much larger observed 
  differences pointed out in the text. Li however may be 
  idiosyncratic rather than being representative of metals 
  generally as suggested by Ref. \onlinecite{stutz99}
\end{references}
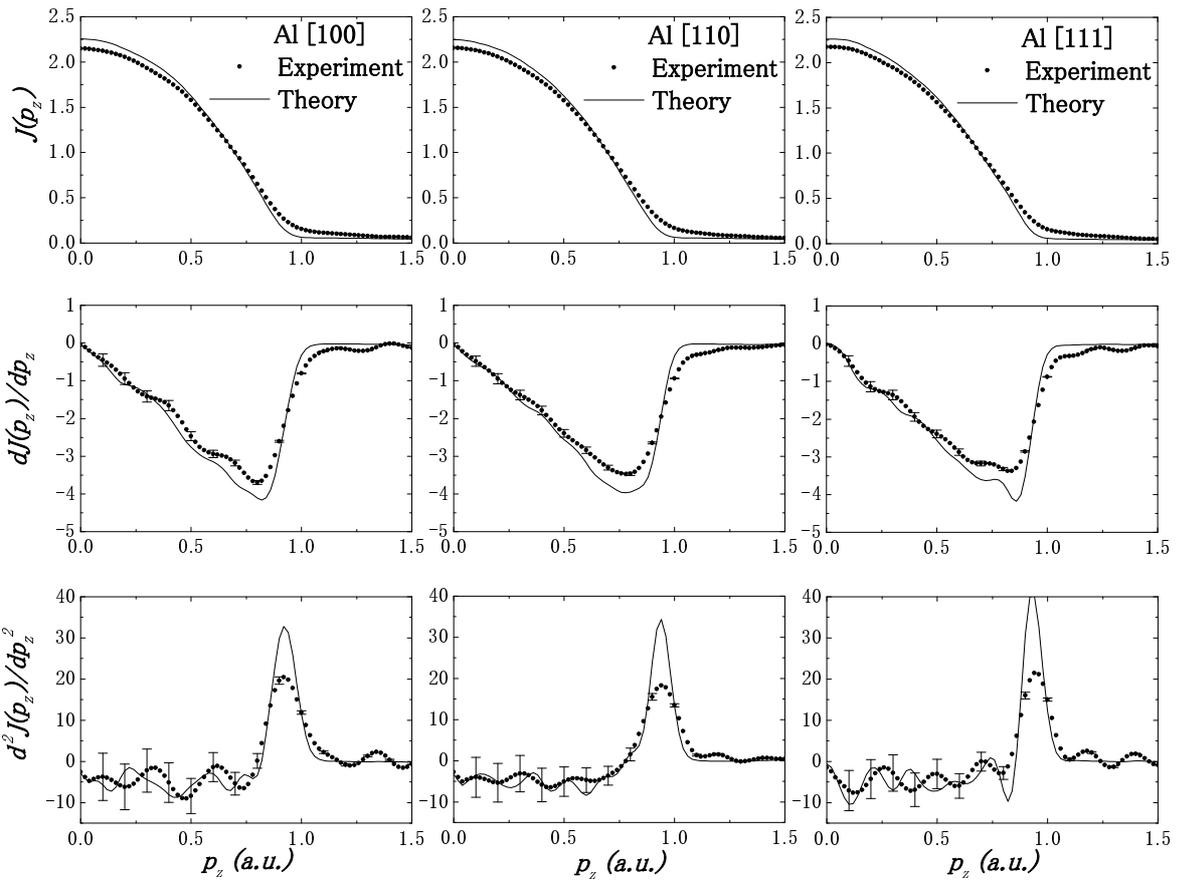
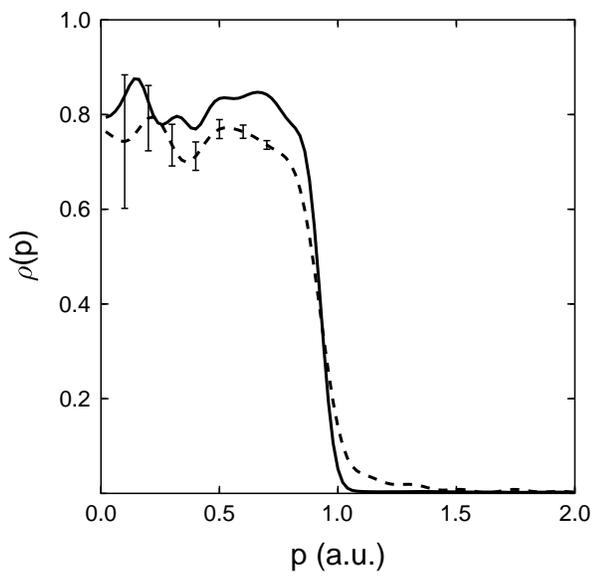
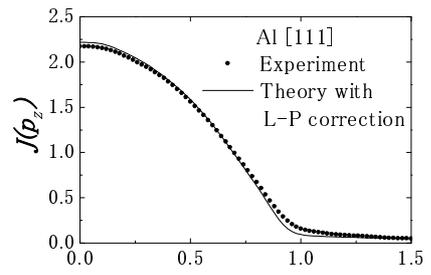
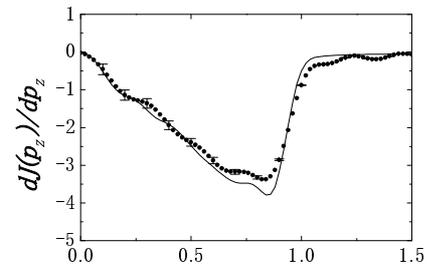
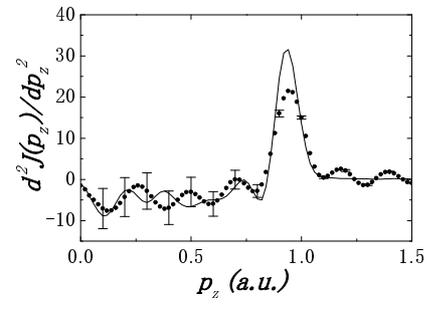
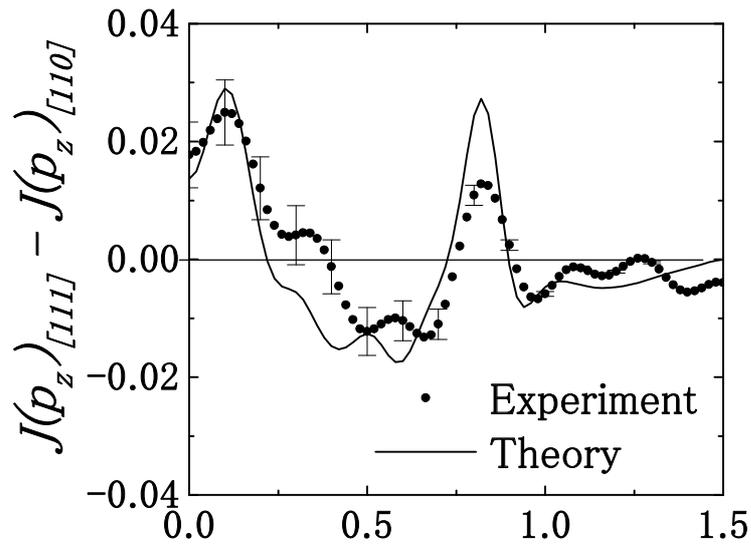
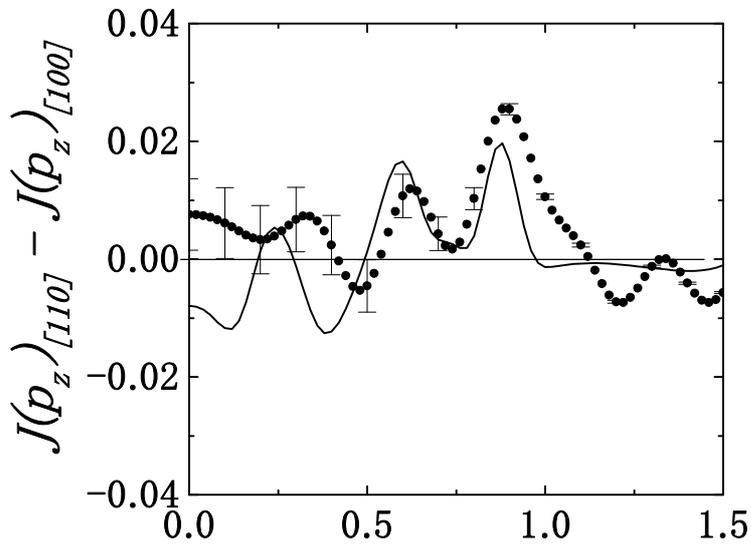
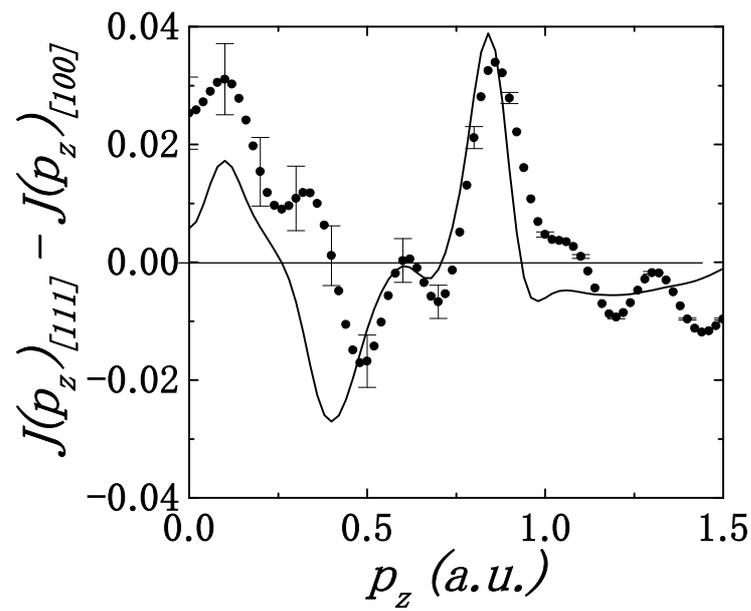
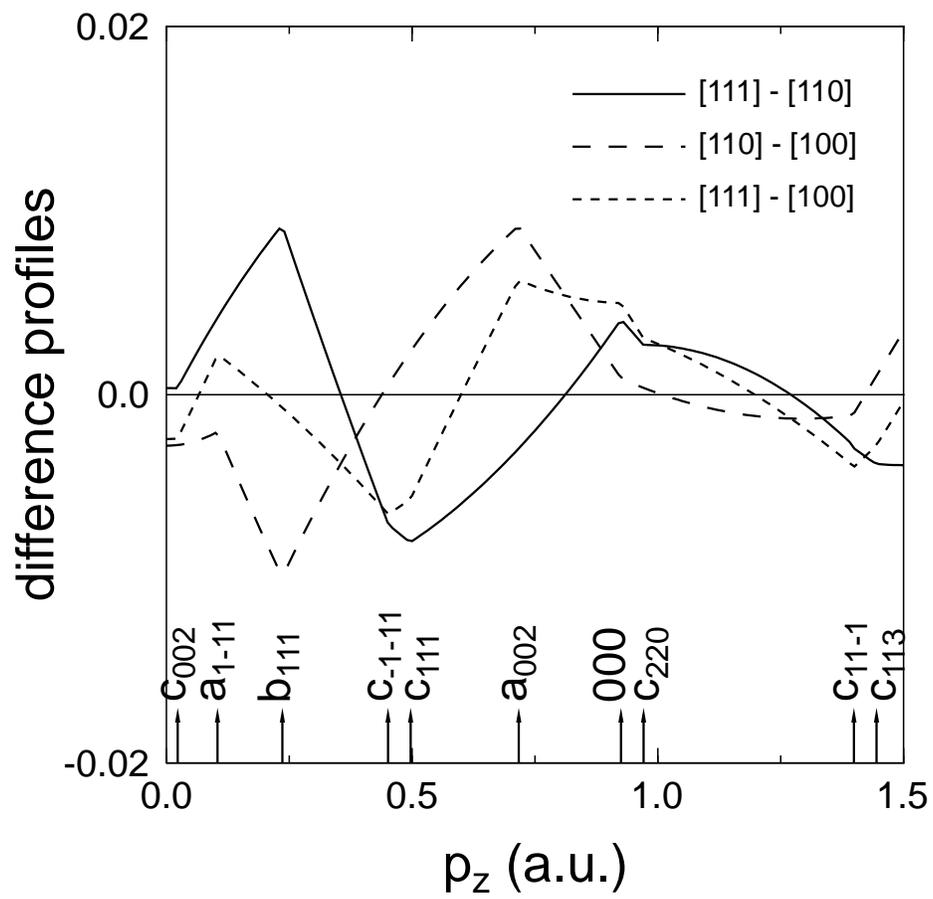
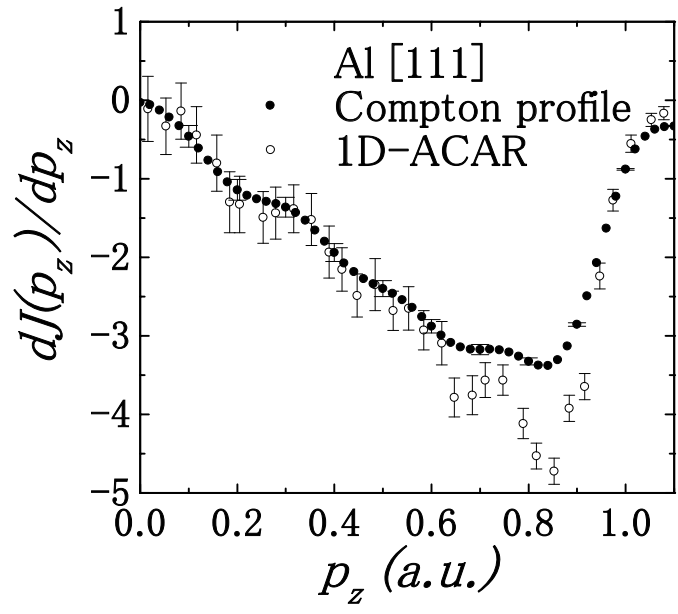
\end{document}